\documentclass[trackchanges,twocolumn]{aastex701}
\usepackage{amsmath}
\usepackage{subcaption}

\newcommand{\br}{\mathbf{r}}

\newcommand{\D}{\mathrm{D}}
\newcommand{\dd}{\mathrm{d}}

\newcommand{\Mach}{\mathcal{M}}

\newcommand{\Rb}{R_{\mathrm{B}}}

\begin{document}

\title{Self-Limited Accretion onto Embedded Binaries in a Uniform Medium}

\author[orcid=0000-0003-3356-880X]{Marcus DuPont}
\affiliation{Department of Astrophysical Sciences, Princeton University, Princeton, NJ, 08544, USA}
\email[show]{marcus.dupont@princeton.edu}  

\author[orcid=0000-0001-9185-5044]{Eliot Quataert} 
\affiliation{Department of Astrophysical Sciences, Princeton University, Princeton, NJ, 08544, USA}
\email{quataert@princeton.edu}  

%% Use the \collaboration command to identify collaborations. This command
%% takes an optional argument that is either a number or the word "all"
%% which tells the compiler how many of the authors above the command to
%% show. For example "\collaboration[all]{(DELVE Collaboration)}" wil include
%% all the authors above this command.
%%
%% Mark off the abstract in the ``abstract'' environment. 
\begin{abstract}
We study accretion from a uniform gas at rest onto equal-mass
binaries -- the binary Bondi problem -- as a function of
adiabatic index~$\gamma$ and compactness $\xi \equiv R_B/a$,
where $R_B$ is the Bondi radius of the binary
and $a$ is the component separation. We present
three-dimensional hydrodynamic simulations spanning
$\xi = \{0.1, 1, 10\}$ at $\gamma = \{1, 4/3, 5/3\}$.
Isothermal gas ($\gamma = 1$) accretes cooperatively at high
compactness, with efficiency $\eta \equiv
\dot{M}_{\rm binary}/\dot{M}_{\rm Bondi} \to 1$ for $\xi \gg
1$ and a stable sonic surface that screens the orbital
modulation. Adiabatic gas ($\gamma > 1$) is self-limiting:
the orbit drives shocks that generate entropy, producing
convective turbulence that suppresses accretion to $\eta
\approx 0.3$ ($\gamma = 4/3$) and $\eta \approx 0.1$
($\gamma = 5/3$), burying the orbital signature in broadband
noise. We derive a stability criterion from first principles:
the sonic surface is the separatrix of the Bondi saddle
point, and the binary annihilates it in $N \propto
(\gamma-1)^{-1}(\sqrt{\xi/\xi_m} - 1)$ orbits, where
$\xi_m = 4/(5{-}3\gamma)$ is the container threshold at
which the sonic surface first encloses the binary, and the
$(\gamma-1)^{-1}$ divergence follows from the lack of entropy generation
at isothermal shocks. For $\gamma = 5/3$, no saddle point exists at any~$\xi$ and
the neutrally stratified Bondi profile is convectively
unstable by a distinct mechanism. The single comparison
$t_{\rm cool}$ versus $NT$ -- where $T$ is the orbital
period -- determines whether an embedded binary accretes
cooperatively or throttles its own fuel supply; simulations
confirm the analytic thresholds and scaling.
\end{abstract}

%% Keywords should appear after the \end{abstract} command. 
%% The AAS Journals now uses Unified Astronomy Thesaurus (UAT) concepts:
%% https://astrothesaurus.org
%% You will be asked to selected these concepts during the submission process
%% but this old "keyword" functionality is maintained in case authors want
%% to include these concepts in their preprints.
%%
%% You can use the \uat command to link your UAT concepts back its source.

%% From the front matter, we move on to the body of the paper.
%% Sections are demarcated by \section and \subsection, respectively.
%% Observe the use of the LaTeX \label
%% command after the \subsection to give a symbolic KEY to the
%% subsection for cross-referencing in a \ref command.
%% You can use LaTeX's \ref and \label commands to keep track of
%% cross-references to sections, equations, tables, and figures.
%% That way, if you change the order of any elements, LaTeX will
%% automatically renumber them.

% \message{The column width is: \the\columnwidth}

\section{Introduction}
\label{sec:intro}

The simplest question about a binary embedded in a uniform gaseous medium is
whether it accretes like a single object. For total mass
$M = m_1 + m_2$, separation~$a$, and ambient sound speed $c_\infty$,
the answer depends on compactness $\xi \equiv R_B/a$, where
$R_B = GM/c_\infty^2$ is the Bondi radius~\citep{Bondi1952}. When
$\xi \ll 1$, each component accretes independently. When $\xi \gg 1$,
the binary is deeply embedded within a shared Bondi sphere and should
accrete cooperatively with efficiency
$\eta \equiv \dot{M}_{\rm binary}/\dot{M}_{\rm Bondi} \to 1$. We
show that this expectation fails for any gas with $\gamma > 1$: the
binary orbit drives shocks that generate entropy, producing convective
turbulence that suppresses accretion and produces sufficient variability in the accretion to suppress the orbital signal. 
%screens the orbital signal.

The interaction of binaries with gas has been studied most extensively
in the circumbinary disk geometry, where angular momentum produces a
rotationally supported structure and the coupling is mediated by
resonant and viscous torques \citep[see e.g.,][]{Artymowicz+Lubow+1994,
Artymowicz+Lubow+1996,MacFadyen+Milosavljevic+2008}; see
\citet{Lai+Munoz+2023} for a review. The complementary regime --
accretion from a uniform, pressure-supported medium at rest onto a
binary potential -- was first simulated by \citet{Farris+2010}, who
performed general-relativistic calculations near coalescence and
coined the term ``binary Bondi.'' This problem was also studied in more detail recently using general-relativistic MHD simulations by \citet{Ressler2025}.  In both works the focus was the electromagnetic
counterpart to merger; at the wide separations relevant to most of a
binary's lifetime, the steady-state thermodynamic response was not
addressed. In the supersonic wind limit, \citet{Antoni+2019} measured
binary Bondi--Hoyle--Lyttleton drag and accretion rates. In the
subsonic limit, linear calculations for single
\citep{Kim+Kim+2007,ONeill+2024} and binary
\citep{Kim+Kim+SanchezSalcedo+2008} perturbers cannot access the
entropy-driven instability we identify, because the linear density
wake is independent of the equation of state. \citet{Kim+2010} entered
the nonlinear regime for a single perturber in $\gamma = 5/3$ gas,
finding hydrostatic envelopes but no convective disruption -- the
quadrupolar forcing of a binary is required.

The binary Bondi problem applies across the electromagnetic and  gravitational wave spectrum. Pulsar timing
arrays have detected the nanohertz background from supermassive black
hole (SMBH) binaries
\citep{NANOGrav+2023,EPTA+2023,Reardon+2023,Xu+2023}, and population
synthesis models often adopt gas hardening prescriptions calibrated to
laminar accretion
\citep{Haiman+2009,Sesana+2013,Kelley+2017a}. LISA
\citep{LISA+2024} will track individual inspiral waveforms through the
regime where gas coupling shapes the orbit. At the stellar-mass end,
LIGO--Virgo--KAGRA
\citep{LIGO+2015,Virgo+2015,KAGRA+2019} detects mergers whose
progenitors may have formed in gaseous environments. Electromagnetic
surveys searching for periodic binary active galactic nuclei (AGN)
signatures have yielded almost uniformly null results:
\citet{ElBadry+2025} tested 181 Gaia DR3 candidates and found all
were false positives. The binary Bondi flow provides a 
mechanism for suppressing periodic accretion signatures and reducing
the effective accretion efficiency of binaries in adiabatic
environments relative to laminar estimates.

In this work, we present three-dimensional inviscid hydrodynamic simulations
spanning $\xi = 0.1$--$10$ at $\gamma = 1$, $4/3$, and $5/3$,
together with an analytic stability criterion that predicts the onset,
timescale, and character of the entropy-driven turbulence.
Section~\ref{sec:theory} derives the criterion.
Section~\ref{sec:methods} describes the numerical setup.
Section~\ref{sec:results} presents results.
Section~\ref{sec:discussion} discusses implications.
Section~\ref{sec:conclusions} summarizes.

%%%%%%%%%%%%%%%%%%%%%%%%%%%%%%%%%%%%%%%%%%%%%%%%%%%%%%%%%%%%%%%%%%%%%%%%%
\section{The Stability Criterion}
\label{sec:theory}

We work in units where $G = \rho_\infty = a = 1$. The mass ratio
is $q \equiv m_2/m_1 \leq 1$ where $M = m_1 + m_2 = 1$.
The binary is on a circular orbit, so its eccentricity $e = 0$.
Three dimensionless parameters characterize the problem: the compactness
$\xi \equiv GM/(c_\infty^2 a) = 1/c_\infty^2$, the adiabatic
index~$\gamma$, and~$q$. The accretion efficiency is
\begin{equation}
    \eta(\xi, \gamma, q)
    \equiv \frac{\dot{M}_{\rm binary}}
                {\dot{M}_{\rm Bondi}(M)}\,,
\end{equation}
where $\dot{M}_{\rm Bondi}(M)$ is the spherical Bondi rate for a
single object of total mass~$M$,
\begin{equation}
    \dot{M}_{\rm Bondi} = 4\pi \lambda(\gamma) \rho_\infty c_\infty \left( \frac{GM}{c_\infty^2} \right)^2 = 4\pi \lambda(\gamma) \xi^{3/2}
    \label{eq"mbondi}
\end{equation}
with $\lambda(\gamma)$ being the accretion coefficient \citep{Bondi1952},
\begin{equation}
    \lambda = \frac{1}{4}\left(\frac{2}{5 - 3\gamma} \right)^{(5 - 3\gamma)/(2\gamma - 2)},
\end{equation}
and $\dot M_{\rm binary}$ is the sum of the accretion rate onto both point masses in the binary problem.

Two scales govern the problem: the
\emph{geometric} compactness at which the accretion flows interact,
and the \emph{thermodynamic} threshold at which that interaction
becomes destabilizing.

\subsection{Geometric Scale}
\label{sec:geometry}

When $\xi \ll 1$, each component accretes independently. Since
$\dot{M}_{\rm Bondi} \propto M^2$, the combined efficiency is
\begin{equation}
    \eta(\xi \ll 1)
    = \frac{1 + q^2}{(1+q)^2}\,,
    \label{eq:eta_independent}
\end{equation}
giving $\eta = 1/2$ for equal masses. The sonic radius of
component~$i$ accreting in isolation is
\begin{equation}
    R_s^{(i)}
    = \frac{5 - 3\gamma}{4}\,\xi\,m_i = R_s\,m_i\,.
    \label{eq:sonic_individual}
\end{equation}
where $R_s = (5 - 3\gamma) \xi/ 4$ is the dimensionless sonic radius for the total mass $M$. Setting $R_s^{(1)} + R_s^{(2)} = 1$ gives the compactness at which
the individual sonic surfaces make geometric contact:
\begin{equation}
    \xi_m(\gamma)
    = \frac{4}{5 - 3\gamma}\,.
    \label{eq:xi_merge}
\end{equation}
This is independent of~$q$, and diverges
as $\gamma \to 5/3$. The compactness $\xi_m$ marks where
the individual accretion flows first interact; whether
that interaction leads to cooperative accretion or to
instability depends on the thermodynamic response of
the gas.

\subsection{Sonic Surface Morphology}
\label{sec:morphology}

Before the sonic surfaces merge, the binary potential imprints a
characteristic geometry on the transonic flow. At $r \gg 1$, the
time-averaged potential of an equal-mass binary ($q = 1$) is
\begin{equation}
    \Phi(r, \theta)
    = -\frac{1}{r}\!\left[
        1 + \frac{1}{8}\!\left(\frac{1}{r}\right)^{\!2}
          P_2(\cos\theta)
        + \mathcal{O}\!\left(\frac{1}{r}\right)^{\!3}
    \right],
    \label{eq:multipole}
\end{equation}
where $\theta$ is the polar angle measured from the orbital axis and $P_2$ is
the Legendre polynomial. 
The quadrupolar perturbation deforms the spherical sonic surface into
$\mathcal{R} = R_s + \delta\mathcal{R}$, with fractional amplitude
\begin{equation}
    \frac{\delta\mathcal{R}}{R_s}
    \sim \frac{1}{8}\!\left(\frac{1}{R_s}\right)^{\!2}
    = \frac{2}{(5 - 3\gamma)^2\,\xi^2}\,,
    \label{eq:quad_amplitude}
\end{equation}
which grows as $\xi$ decreases or $\gamma \to 5/3$. The perturbation
rotates at $\Omega = 1$, producing an $m = 2$ spiral pattern in the
corotating frame. \citet{Foglizzo+Ruffert+1997} proved that any
detached sonic surface must intersect $R_s$ at least once; in the
binary case this constrains the deformation to two trailing spiral
arms connecting the individual Bondi spheres to the shared transonic
flow.

For $\gamma = 5/3$, $R_s = 0$ identically: no detached sonic surface
exists at any~$\xi$. The flow is always of Type~SF or FSF in the
\citet{Foglizzo+Ruffert+1997} classification -- the sonic surface is
attached to the accretors and directly exposed to the orbital
dynamics. Whether these spirals remain stationary or become unstable
depends on the entropy they carry. 

\subsection{Thermodynamic Scale: Shock-Driven Instability}
\label{sec:entropy_source}

The binary orbit introduces irreversible entropy generation
absent in single-body Bondi accretion. Whether this entropy
destabilizes the transonic flow depends on three factors: the
rate of generation, the non-axisymmetry of the source, and
whether the sonic sphere can confine the heated gas.

\subsubsection{Propeller Mach number}

The gravitational potential of the binary is a quadrupolar
pattern rotating at $\Omega = 1$ in the inertial frame. In
the corotating frame, this potential is time-independent. Gas
falling from infinity with zero angular momentum has
$v_\phi = 0$ in the inertial frame; transforming to the
corotating frame gives an azimuthal velocity
$v_\phi' = -\Omega\,r\sin\theta$ at position $(r,\theta)$,
where $\theta$ is the polar angle measured from the orbital
axis. The gas therefore streams through the static corotating
potential at azimuthal speed $\Omega\,r\sin\theta$.

The sound speed at $r = 1$ follows from adiabatic
compression during infall. In the supersonic interior ($r < R_s$), 
mass conservation in the freefall regime gives
$\rho \propto r^{-3/2}$, so that
$c^2 \propto \rho^{\,\gamma-1} \propto r^{-3(\gamma-1)/2}$.
Evaluating at $r = 1$ and normalizing to $c_\infty$ at the
Bondi radius gives
$c(1) = c_\infty\,\xi^{3(\gamma-1)/4}$. (For
$\xi > \xi_m$ the binary sits inside the sonic surface and
the freefall scaling applies; for $\gamma = 5/3$, the flow is
everywhere marginally transonic and $\rho \propto r^{-3/2}$
is exact.) The propeller Mach number -- the ratio of the
corotating azimuthal speed to the local sound speed -- is
therefore
\begin{equation}
    \mathcal{M}_p(\theta)
    \equiv \frac{|\,v_\phi'\,|}{c(1)}
    = \xi^{(5-3\gamma)/4}\sin\theta\,,
    \label{eq:mach_prop}
\end{equation}
which we evaluate at its maximum, $\theta = \pi/2$ (the
orbital plane):
$\mathcal{M}_p \equiv \xi^{(5-3\gamma)/4}$.
For $\gamma < 5/3$ and $\xi > 1$, $\mathcal{M}_p > 1$: the
gas sweeps through the quadrupolar pattern supersonically.
Shocks form only where
$\mathcal{M}_p\sin\theta > 1$, i.e.,\ outside a polar cone
of half-angle $\theta_c = \arcsin(\mathcal{M}_p^{-1})$. For deeply
embedded binaries ($\mathcal{M}_p \gg 1$), $\theta_c \to 0$
and nearly the full solid angle is shocked; at marginal
compactness ($\mathcal{M}_p \sim 2$), $\theta_c \approx
30^\circ$ and gas accreting through the polar funnels arrives
unshocked. The entropy generation is therefore concentrated
in the equatorial belt.

The non-axisymmetric component of the potential at $r = 1$ is
dominated by the quadrupole (Eq.~\ref{eq:multipole}), with
amplitude $\delta\Phi \sim q(1+q)^{-2}$, yielding velocity
perturbations $\delta v \sim \delta\Phi/v_{\rm orb} 
\sim q(1+q)^{-2}$\ from balancing the advective 
and potential gradient terms in the linearized momentum  
equation in the supersonic limit, $\mathcal{M}_p > 1$.
For equal masses, $\delta v \approx 1/4$ -- a
finite-amplitude perturbation. Compressive phases steepen
into shocks on the nonlinear acoustic timescale
$\tau_{\rm steep} \sim c/[(\gamma+1)\,\omega\,\delta v]$
\citep[\S101 in][]{Landau+Lifshitz+1959}, with
$\omega = 2\Omega$ (the gas encounters two compression
maxima per azimuthal circuit through the $m = 2$ pattern).
This gives
\begin{equation}
    \Omega\,\tau_{\rm steep}
    \sim \frac{(1+q)^2}
             {2q\,(\gamma+1)\,\mathcal{M}_p}\,,
    \label{eq:tau_steep}
\end{equation}
the shock formation time in units of $\Omega^{-1}$. For
$q = 1$ and $\mathcal{M}_p \gtrsim 1$,
$\Omega\,\tau_{\rm steep} \lesssim 1$: standing shocks form
within a single orbital period in the corotating frame. For
$\mathcal{M}_p < 1$ the streaming is subsonic and the gas
adjusts quasi-statically through smooth pressure gradients;
the flow remains isentropic. In the unperturbed Bondi profile, $\mathcal{M}_p = 1$
identically for $\gamma = 5/3$: nonlinear steepening of
the background flow is arrested at the marginally sonic
threshold. The binary's finite-amplitude perturbation
creates local regions where $\mathcal{M}_p$ exceeds
unity, but no global standing shock pattern forms in the co-rotating frame.

\subsubsection{Entropy generation}

The dimensionless entropy jump across a shock (Eq.~1.85 in
\citealt{Zeldovich+Raizer+1966}):
\begin{equation}
    \Delta s = \ln \left[\frac{2\gamma \mathcal{M}_1^2 - (\gamma - 1)}{\gamma + 1} \right] - \gamma \ln \left[\frac{(\gamma + 1)\mathcal{M}_1^2 }{(\gamma - 1)\mathcal{M}_1^2 + 2} \right]
    \label{eq:s_jump}
\end{equation}
vanishes identically at $\gamma = 1$ for all Mach numbers $\mathcal{M}_1$ where the subscript represents the upstream value. Expanding about this exact zero yields
\begin{equation}
    \Delta s
    = (\gamma - 1)\,g(\mathcal{M}_1)
    + \mathcal{O}\!\left((\gamma-1)^2\right),
    \label{eq:entropy_leading}
\end{equation}
where
$g(\mathcal{M}) \equiv (\mathcal{M}^4 - 1)/(2\mathcal{M}^2)
- \ln\mathcal{M}^2$
is positive-definite for $\mathcal{M} > 1$. This is the leading term in an
expansion about the isothermal zero, valid for shocks of
arbitrary strength.

The upstream Mach number $\mathcal{M}_1$ entering the
Rankine--Hugoniot relation is the component of velocity
\emph{normal} to the shock front, measured in the shock
frame. In the corotating frame, the gas streams through the
static quadrupolar potential at azimuthal speed
$\mathcal{M}_p\,c(1)$. The standing shocks that form from
the steepened perturbation are generically oblique: gas hits
the shock at angle $\beta$ to the shock normal, so
$\mathcal{M}_1 = \mathcal{M}_p\sin\beta \leq \mathcal{M}_p$.
The identification $\mathcal{M}_1 \sim \mathcal{M}_p$ is
therefore an upper bound on the shock strength; the true
entropy generation per shock passage is reduced by an
order-unity geometric factor that depends on the shock
morphology. The essential scaling $\Delta s \propto (\gamma -
1)$, however, is exact: it follows from the equation of state
and is independent of the shock geometry.

A uniform entropy offset merely shifts the Bondi solution to a
new baseline; it does not destabilize the flow. What breaks the
sonic point is an entropy \emph{gradient} across streamlines.
The binary potential is quadrupolar: streamlines at different
azimuths encounter shocks of different strength and obliquity.
The resulting entropy variation scales as
\begin{equation}
    \delta s
    \sim (\gamma - 1)\,
      f(\mathcal{M}_p)\,
      \frac{q}{(1+q)^2}\,,
    \label{eq:entropy_var}
\end{equation}
where $f(\mathcal{M}_p)$ is a positive-definite function that
vanishes at $\mathcal{M}_p = 1$ and grows for
$\mathcal{M}_p > 1$; its precise form depends on the shock
geometry, but its existence requires only that different
streamlines experience different compressions. For $\gamma =
1$ the variation vanishes (no fuel); for $\mathcal{M}_p = 1$
it vanishes (no shock); for $q = 0$ it vanishes (no
asymmetry). This entropy gradient violates the Schwarzschild
stability criterion ($\dd s/\dd r < 0$), seeding convective
instability.

% \subsubsection{The container condition}
\subsubsection{Annihilating the Saddle Point}
Entropy generated at $r \sim 1$ can only accumulate if the
sonic surface encloses the binary. For $\xi \geq \xi_m$,
the binary sits inside the sonic surface: acoustic signals
cannot propagate outward through the supersonic infall, and
the shocked gas is confined. Each orbit adds entropy to
the trapped region. For $\xi < \xi_m$, the binary is in
the subsonic exterior and acoustic perturbations disperse
freely. Whether the accumulated entropy disrupts the
transonic structure -- and on what timescale -- is
determined by the energy budget.

The sonic surface is the separatrix of the Bondi saddle 
point. In the $(r, v)$ phase plane, the critical point at 
$R_s$ is an X-point: the transonic accretion branch and the 
transonic wind branch cross there, and every other 
trajectory is either purely subsonic or purely supersonic. 
The transonic accretion solution is the unique trajectory 
threading the saddle, selected by the eigenvalue at the 
critical surface. The global accretion rate is determined 
there, and only there.

Each binary orbit deposits an entropy increment $\delta s 
\sim (\gamma-1)f(\mathcal{M}_p)\,q/(1+q)^2$ into the 
shocked gas at $r \sim 1$ (Eq.~\ref{eq:entropy_var}), 
generating a fractional pressure perturbation $\delta P / P 
= \delta s$ at fixed density. The X-point at $R_s$ is 
indifferent to conditions downstream of it: as long as the 
saddle persists, the transonic topology is intact and the 
global accretion rate is still selected by the critical 
surface. The convective plumes born at $r \sim 1$ must 
propagate outward through the entire supersonic shell and 
deliver enough entropy to $R_s$ to eliminate the X-point. 
The disruption condition is annihilation of the saddle 
itself.

Given the container condition $\xi > \xi_m$ 
(Section~\ref{sec:geometry}), the sonic surface encloses 
the binary and the supersonic shell between $r = 1$ and $r 
= R_s$ is well defined. The mechanical energy cost of 
transporting the perturbation from source to separatrix, 
integrated over this shell, is
\begin{equation}
    E_{\rm disrupt} 
    \sim \dot{M}_{\rm Bondi}
         \left(\sqrt{\frac{\xi}{\xi_m}} - 1\right)\,.
    \label{eq:Ebreak}
\end{equation}
This is a first-order estimate valid for $\xi > \xi_m$. 
At $\xi = \xi_m$, the source and the separatrix coincide 
and $E_{\rm disrupt} = 0$: entropy is deposited directly 
at the X-point and no transport is required. For $\xi \gg 
\xi_m$, $E_{\rm disrupt} \sim \dot{M}_{\rm Bondi}\,
\sqrt{\xi/\xi_m}$ grows with embedding because the shell 
between source and saddle grows. The quadrupolar 
perturbation at $r = 1$ is undiminished -- its amplitude 
is fixed by $q/(1+q)^2$ and independent of $\xi$ -- but 
the binary looks increasingly like a monopole as seen from 
$R_s$, since $a/R_s \propto 1/\xi \to 0$. The 
$\sqrt{\xi/\xi_m}$ factor measures this transport cost.

The binary deposits energy via gravitational stirring at a 
rate $\dot{E}_{\rm stir} = \zeta\,\dot{M}_{\rm Bondi}$, 
where $\zeta$ absorbs the entropy generation efficiency 
per shock, the retention fraction against advective losses 
into the sinks, and the acoustic coupling efficiency 
between $r \sim 1$ and $R_s$. All three factors satisfy 
$\zeta \propto (\gamma - 1)$: the exact Rankine--Hugoniot 
relation (Eq.~\ref{eq:s_jump}) requires isothermal shocks 
to generate zero entropy, making the stirring 
thermodynamically inert regardless of Mach number. Over 
one orbital period $\Delta E_{\rm orb} = 2\pi\zeta\,
\dot{M}_{\rm Bondi}$, and the number of orbits to 
annihilate the saddle is
\begin{equation}
    N = \frac{E_{\rm disrupt}}{\Delta E_{\rm orb}}
      \sim \frac{1}{\zeta}
           \left(\sqrt{\frac{\xi}{\xi_m}} - 1\right)
      \propto \frac{1}{\gamma - 1}
              \left(\sqrt{\frac{\xi}{\xi_m}} - 1\right)\,,
    \label{eq:N_orbits}
\end{equation}
where the magnitude of $\zeta$ is not predicted from first 
principles. 
% The observed onset of turbulence within 
% $\mathcal{O}(1)$--$\mathcal{O}(10)$ orbits at $\xi = 
% \xi_m$ for $\gamma = 4/3$ constrains $\zeta \sim 
% 0.1$--$1$: at threshold $\sqrt{\xi/\xi_m} - 1 = 0$ and 
% $N$ is dominated entirely by $\zeta$, consistent with the 
% rapid onset found numerically.

\subsubsection{Limiting cases}

As $\gamma \to 1$, $N \to \infty$ through $\zeta \to 0$:
isothermal shocks deposit zero entropy by 
Eq.~\eqref{eq:s_jump}, the plumes carry no buoyancy, and 
the saddle point is never reached regardless of how long 
the binary orbits. The gas has infinite specific heat; 
shock heating generates no usable entropy and the stirring 
is thermodynamically inert.

As $\gamma \to 5/3$, $\xi_m \to \infty$ 
(Section~\ref{sec:geometry}) and Eq.~\eqref{eq:N_orbits} 
never applies. In the freefall regime, both the infall 
velocity and the sound speed scale as $r^{-1/2}$; their 
ratio is frozen and the flow is everywhere marginally 
transonic. No X-point exists at any finite radius, so 
there is no separatrix to protect the flow. The propeller 
shocks vanish ($\mathcal{M}_p = 1$ generates no entropy 
jump), but the flow is nonetheless unstable: the binary 
potential creates finite-amplitude perturbations to the 
density and velocity fields, and in a $\gamma = 5/3$ gas 
-- where the adiabatic Bondi profile is neutrally 
stratified -- any such perturbation produces entropy 
inversions that violate the Schwarzschild criterion 
($\mathrm{d}s/\mathrm{d}r < 0$), seeding convective 
instability. Turbulence at $\gamma = 5/3$ is therefore 
thermodynamic in origin rather than shock-driven: it 
arises from entropy \emph{rearrangement} by the binary 
potential in a flow with zero convective stability margin. 
The resulting turbulence is nonetheless shock-dominated in 
its saturated state: the convective velocities are 
transonic by construction ($v/c \lesssim 1$ at all radii 
for $\gamma = 5/3$), so the turbulent eddies themselves 
steepen into shocks (Section~\ref{sec:turbulence_character}).

As $\xi \to \infty$ ($a \to 0$ at fixed $M$ and 
$c_\infty$), $N \to \infty$ through the growing transport 
barrier $\sqrt{\xi/\xi_m}$: the shell between source and 
separatrix grows without bound, recovering the known 
stability of single Bondi accretion \citep{Bondi1952}.
The entropy generated per orbit involves a competition
between fuel ($\gamma - 1$, increasing toward $5/3$) and 
propeller strength ($\mathcal{M}_p$, decreasing toward 
$5/3$), which peaks near $\gamma \sim 1.2$--$1.4$: the 
relativistic equation of state $\gamma = 4/3$ sits in the 
most potent range.

We define breakout as the annihilation of the Bondi saddle 
point: the moment when cumulative entropy injection 
eliminates the X-point in the $(r,v)$ phase plane and the 
transonic accretion solution ceases to exist as a smooth 
trajectory. Breakout of the convective region through the 
large-scale sonic surface requires three ingredients:
(i)~a \emph{propeller} ($\xi > 1$, $\gamma < 5/3$): the
pattern moves supersonically and generates shocks;
(ii)~\emph{fuel} ($\gamma > 1$): the post-shock gas retains
entropy and develops convectively unstable gradients; and
(iii)~a \emph{container} ($\xi > \xi_m$): a finite sonic
surface confines the heated bubble. When all three are
satisfied, breakout is inevitable in the absence of radiative
cooling and occurs on the timescale given by
Eq.~\eqref{eq:N_orbits}. Cooling introduces a fourth condition: the cooling time at
$r \sim 1$ must exceed $N\,T$, where $T$ is the orbital
period. When $t_{\rm cool} \ll NT$, post-shock entropy is
radiated before it accumulates, and the gas behaves as
effectively isothermal regardless of its microscopic
$\gamma$. The single comparison $t_{\rm cool}$ versus $NT$
therefore governs whether a given astrophysical system
falls in the cooperative or turbulent regime.
%%%%%%%%%%%%%%%%%%%%%%%%%%%%%%%%%%%%%%%%%%%%%%%%%%%%%%%%%%%%%%%%%%%%%%%%%%%%%%%%%%
\section{Numerical Methodology}
\label{sec:methods}

The problem requires resolving the interaction between a global sonic
surface at $r \sim \xi$ and the vacuum-like accretion funnels at
$r \sim 1$, spanning Mach numbers from $\Mach \sim 0$ at stagnation
points to $\Mach \gg 1$ in the free-fall region. We carry out this
numerical experiment using the open-source Godunov-type gas dynamics
code \texttt{SIMBI}~\citep{Dupont+2023}.

\subsection{Governing Equations}

We solve the Newtonian continuity and Euler equations,
%%%%%
\begin{eqnarray}
    \frac{\partial \rho}{\partial t} + \nabla \cdot (\rho \mathbf{v}) &=& S_\rho\label{eq: den}, \\
    \frac{\partial (\rho \mathbf{v})}{\partial t} + \nabla \cdot (\rho \mathbf{v} \otimes \mathbf{v} + P \mathbb{I}) &=& -\rho \nabla \Phi + \mathbf{S}_{\rho \mathbf{v}}\label{eq: mom}, \\
    \frac{\partial \mathcal{E}}{\partial t} + \nabla \cdot [(\mathcal{E} + P)\mathbf{v}] &=& -\rho \mathbf{v} \cdot \nabla \Phi + S_\mathcal{E}\label{eq: nrg},
\end{eqnarray}
%%%%%%%
where $\rho$ is the fluid density, $\mathbf{v}$ the velocity field,
$P$ the pressure, and
$\mathcal{E} = \frac{1}{2}\rho v^2 + \epsilon_{\rm int}$ the total
energy density with $\epsilon_{\rm int}$ being the internal energy density. The source terms
$S_\rho, \mathbf{S}_{\rho \mathbf{v}}, S_\mathcal{E}$ represent
explicit removal of mass, momentum, and energy by the point
masses~\citep{Krumholz+2004}. The gravitational potential is
\begin{equation}
    \Phi(\br, t) = -\sum_{i=1}^{2} \frac{M_i}{\sqrt{|\br - \br_i(t)|^2 + \ell_i^2}},
\end{equation}
where $\br_i(t)$ is the position of mass $i$ and the softening length
$\ell_i$ is fixed at 1\% of the individual Bondi radius. We close the
system with a gamma-law equation of state,
$P = (\gamma - 1)\epsilon_{\rm int}$, and investigate two
thermodynamic limits: isothermal ($\gamma = 1$) to isolate the pure
gas-dynamic response to the binary potential, and adiabatic
($\gamma > 1$) to explore the role of shock heating. For the
isothermal case, uniform initial conditions ($\rho_\infty$,
$c_\infty$) are numerically stable; the adiabatic runs required mild energy floors at start up ($\epsilon_{\rm floor} = 10^{-6}$ in code units, six orders
of magnitude below the ambient thermal energy density) for
$\xi = 10$.

We evolve Equations~\ref{eq: den}--\ref{eq: nrg} with the HLLE
Riemann solver~\citep{HLL+1983, Einfeldt+1988} and piecewise linear
reconstruction (Minmod, $\theta_{\rm plm} = 1$) across active zones and piecewise parabolic reconstruction across level boundaries to cancel out the order of spatial accuracy loss \citep{Berger+Colella+1989}. The HLLE solver is
chosen for its positivity-preserving property -- essential in the
evacuated regions between accretors -- and because its numerical
dissipation at contact discontinuities suppresses the persistent
oscillations that contact-resolving schemes (e.g., HLLC, Roe) excite at the
standing shocks characteristic of this problem.

\subsection{The Telescoping Grid}

To bridge the scale separation of $\xi \sim 10$ without prohibitive
cost, we implement a telescoping grid using fixed mesh refinement
(FMR), a static variant of the adaptive mesh refinement framework of
\cite{Berger+Colella+1989}. The grid hierarchy is a sequence of
nested Cartesian patches centered on the origin. The base grid covers
$\pm R_{\rm out} = \texttt{max}(4\,\Rb, 4a)$, ensuring the outer boundary is causally
disconnected from the binary's acoustic feedback for the duration of
each simulation. Each successive level halves the semi-width and
doubles the resolution, maintaining a constant zone count per patch.
The finest level covers the orbital domain ($r \le 1.5$) at
$\Delta x_{\rm fine} = 1/64$, decreased to $1/256$ for $\xi = 0.1$
to place at least 5--10 zones inside the sonic surface or Bondi surface for the runs where there is no sonic surface.

\begin{figure*}[t]
    \centering
    \includegraphics[width=\textwidth]{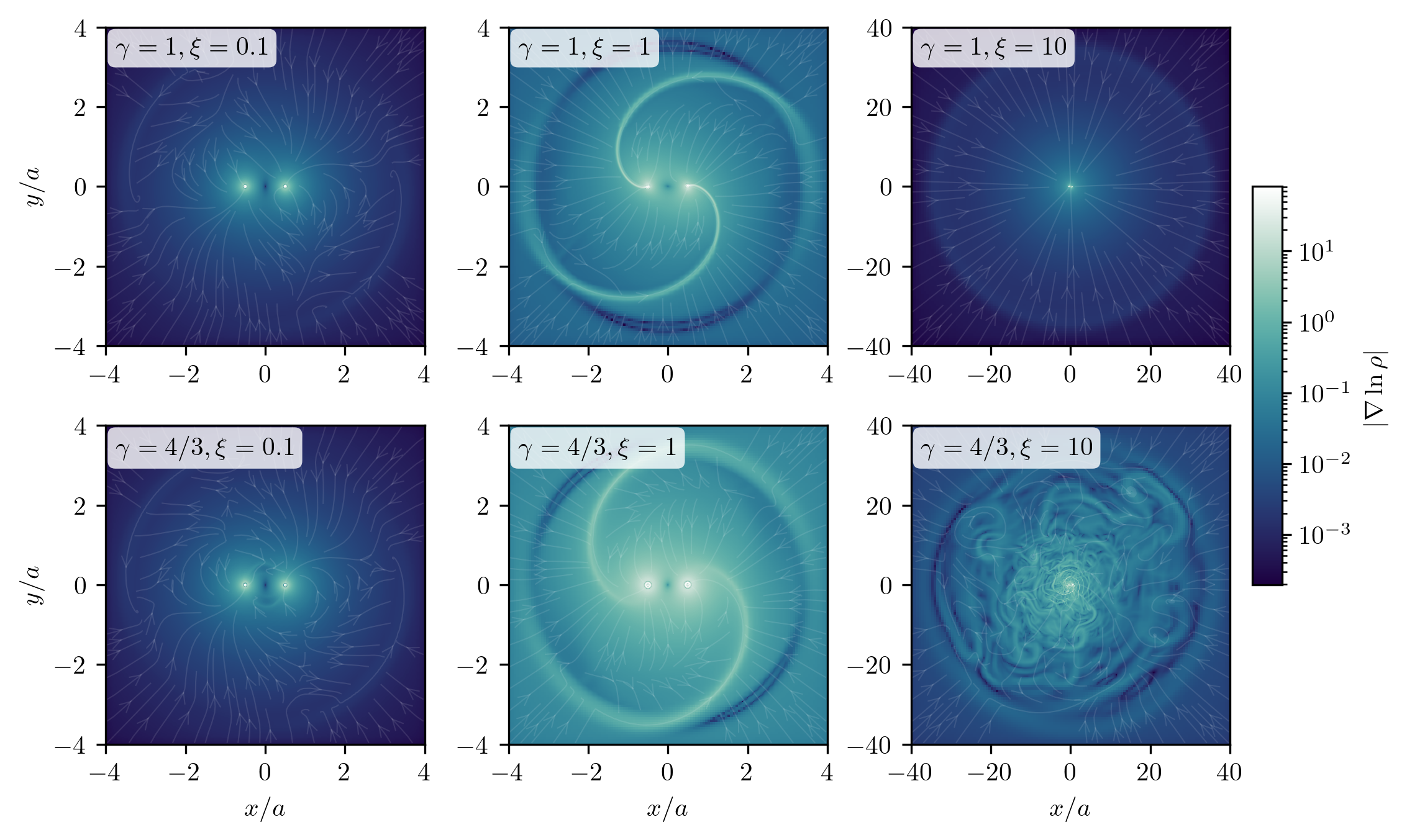}
    \caption{Density gradient schlieren $|\nabla \ln \rho|$ with velocity 
    streamlines overlaid, showing flow morphology across compactness $\xi$ 
    for isothermal ($\gamma=1$, top) and relativistic ($\gamma=4/3$, bottom) 
    gas. Isothermal flows remain stationary and organized at all $\xi$: 
    independent accretion at $\xi=0.1$, coherent spiral structure at $\xi=1$, 
    and radial monopole-like inflow at $\xi=10$. Polytropic flows exhibit 
    similar organization at low $\xi$ but become turbulent at high $\xi$: 
    $\xi=10$ shows complete loss of coherent structure with chaotic, 
    multi-scale density fluctuations. The transition from organized to 
    turbulent occurs between $\xi=1$ and $\xi=10$ for $\gamma=4/3$, 
    consistent with the predicted threshold Eq.~\eqref{eq:xi_merge} where shock-generated entropy overwhelms the sonic buffer. All above plots are shown at $t \geq 3t_B$ for each respective $\xi$.}
    \label{fig:iso_rel_schlieren}
\end{figure*}
\subsection{Convergence}
\label{sec:convergence}

The accretion efficiency $\eta = \dot{M}_{\rm binary}/\dot{M}_{\rm Bondi}$
is a ratio of two quantities computed with identical sink prescriptions:
the same removal radius $r_{\rm acc}$, the same softening length $\ell$,
and the same energy floor. Any systematic bias introduced by the sink
enters both numerator and denominator and cancels in the ratio. This
self-calibration is the principal reason $\eta$ is the appropriate
accretion metric.

Two dimensionless ratios control the residual sensitivity to the sink.
The first is $r_{\rm acc}/R_s$: the sink must lie well inside the sonic
surface so that the transonic structure that sets $\dot{M}_{\rm Bondi}$
is resolved. We find that the single-accretor rate converges to the
analytic Bondi solution within $\lesssim 3\%$ once
$r_{\rm acc}/R_s \lesssim 0.1$, consistent with the resolution
requirements identified by \citet{Ruffert+1994} and
\citet{Krumholz+2004}. The second is $r_{\rm acc}/a$: the sink must be
small compared to the binary separation so that the two accretion
funnels are resolved as distinct objects. With $r_{\rm acc}/R_s$
held fixed at a value satisfying the first condition, we vary
$r_{\rm acc}/a$ over a decade ($0.10$, $0.05$, $0.01$) and find that
the time-averaged $\eta$ in statistical steady state changes by less
than $5\%$. The softening length $\ell/a$ was tested independently
at $0.10$, $0.05$, and $0.01$ with no significant change in $\eta$.
The insensitivity of $\eta$ to both $r_{\rm acc}/a$ and $\ell/a$
confirms that the accretion suppression is governed by the
thermodynamic structure at $r \sim a$--$R_s$ -- the entropy bubble
and the sonic surface -- rather than by the details of mass removal
near each component.

For $\gamma = 5/3$, the Bondi solution has no detached sonic point:
$R_s = 0$ identically, and the condition $r_{\rm acc}/R_s \ll 1$ is
undefined. The absolute accretion rate retains a residual dependence
on $r_{\rm acc}$, as noted by \citet{Ruffert+1994}. However, both
$\dot{M}_{\rm binary}$ and $\dot{M}_{\rm single}$ are computed with
the same $r_{\rm acc}$, and the $r_{\rm acc}$-dependence cancels in
$\eta$. We verify this explicitly: $\eta(\xi)$ at $\gamma = 5/3$ is
unchanged when $r_{\rm acc}/a$ is varied over the same decade. This
is the stronger convergence statement -- neither numerator nor
denominator converges individually, but their ratio does -- and it
confirms that $\eta$ measures a property of the binary's
thermodynamic interaction with the gas. At $\xi = 0.1$, the softening length $\ell$ falls below the grid scale on the finest level; the insensitivity of $\eta$ to $\ell/a$ over a decade confirms that this does not affect
the results.

\subsection{Boundary Conditions}
\label{sec:boundary}
We employ a relaxation boundary condition in the outer 20\% of the
domain ($R_{\rm buf} > 0.8\,R_{\rm out}$), applying a source term
that drives the flow toward the analytic Bondi profile for a monopole
of mass $M$:
\begin{equation}
    \frac{\partial U}{\partial t} = \dots - \frac{1}{\tau_{\rm damp}} \mathcal{S}(r) \left( U - U_{\rm Bondi} \right),
\end{equation}
where $\mathcal{S}(r)$ is a cubic smoothstep with values in $[0,1]$,
$U$ the conserved state, $U_{\rm Bondi}$ the target Bondi state, and
$\tau_{\rm damp}$ the lesser of the sound-crossing time across the
buffer zone and half the orbital period. This ensures the asymptotic
far-field mass flux,
%%%%%
\begin{equation}\label{eq: bondi_buffer}
    \dot{M}_{\rm buf}(r)\approx 4\pi \lambda(\gamma) \xi^{3/2} \left[1 - \frac{1}{4}\left(\frac{\xi}{r}\right)^2\right],
\end{equation}
%%%%%
is obeyed, with $\lambda(\gamma)$ being the eigenvalue from the Bondi solution. This active boundary ensures that the mass supply is determined by the cloud thermodynamics rather than by artificial reflection or rarefaction waves. 
Relaxing the full conserved state (density, momentum,
and energy) is essential. With standard outflow
boundary conditions, the domain depletes faster than gas
can be resupplied: the accretion rate falls below the
theoretical Bondi value even for a single accretor, as
the finite mass reservoir is exhausted before a steady
state is reached. The relaxation boundary continuously
replenishes the outer domain toward the analytic profile,
ensuring an effectively infinite mass supply and allowing
the flow to settle to its true steady state. We emphasize that 
this active boundary controls the \emph{supply} of gas at 
$r \sim R_{\rm out}$, not the accretion rate onto the point masses; 
the interior flow is free to find whatever steady state the local 
thermodynamics demands.

%%%%%%%%%%%%%%%%%%%%%%%%%%%%%%%%%%%%%%%%%%%%%%%%%%%%%%%%%%%%%%%%%%%%%%%%%%%%%%%%%%
\section{Results}
\label{sec:results}

All simulations are evolved for at least $3\,t_B$
after transients decay where $t_B = R_B / c_\infty = \xi^{3/2}$ is the characteristic Bondi time. Statistics are computed over 5-orbit
windows for $t > t_B$. We present the flow
morphology, turbulence diagnostics, temporal variability, and
accretion efficiency in turn.

\subsection{Flow Morphology}
\label{sec:morphology_results}

Figure~\ref{fig:iso_rel_schlieren} shows density gradient
schlieren $|\nabla \ln\rho|$ with velocity streamlines for
$\gamma = 1$ (top) and $\gamma = 4/3$ (bottom) across
$\xi = 0.1$, $1$, and $10$. The smooth transitions visible 
at large radii in each figure is the imprint of the 
relaxation buffer (Section~\ref{sec:boundary}). The flow 
is driven toward the analytic Bondi profile in the outer 
20\% of the domain, producing a featureless gradient there 
by construction. All physical structures discussed below 
lie well inside this buffer region.

For isothermal gas, the flow is stationary in the corotating
frame at all~$\xi$. At $\xi = 0.1$ the components accrete
independently, each with its own Bondi sphere. At $\xi = 1$
the sonic surfaces interact, producing a coherent $m = 2$
spiral shock pattern. At $\xi = 10$ the spiral arms have
merged into a high-density bridge connecting the two
components, and the outer flow is indistinguishable from
radial monopole infall. The bridge is the mechanism by which $\eta \to 1$: it
channels mass directly to the individual accretion funnels
through a shared high-density stream. 
{Figure~\ref{fig:iso_rel_zoom} shows the Mach number 
field $\mathcal{M} = |\mathbf{v}|/c$ and density gradient 
schlieren at $\xi = 10$ for both $\gamma = 1$ and $\gamma 
= 4/3$. For isothermal gas the sonic surface 
($\mathcal{M} = 1$) is smooth and circular -- the Bondi 
saddle point is intact -- and the bridge connecting the 
two accretors is laminar with no shocks outside the 
immediate vicinity of the accretors.
% Figure~\ref{fig:iso_rel_zoom} shows the orbital-scale structure 
% at $\xi = 10$ in detail: the isothermal bridge
% is smooth and laminar, with no shocks outside the
% immediate vicinity of the accretors.
%
\begin{figure*}[t]
    \centering
    \includegraphics[width=\textwidth]{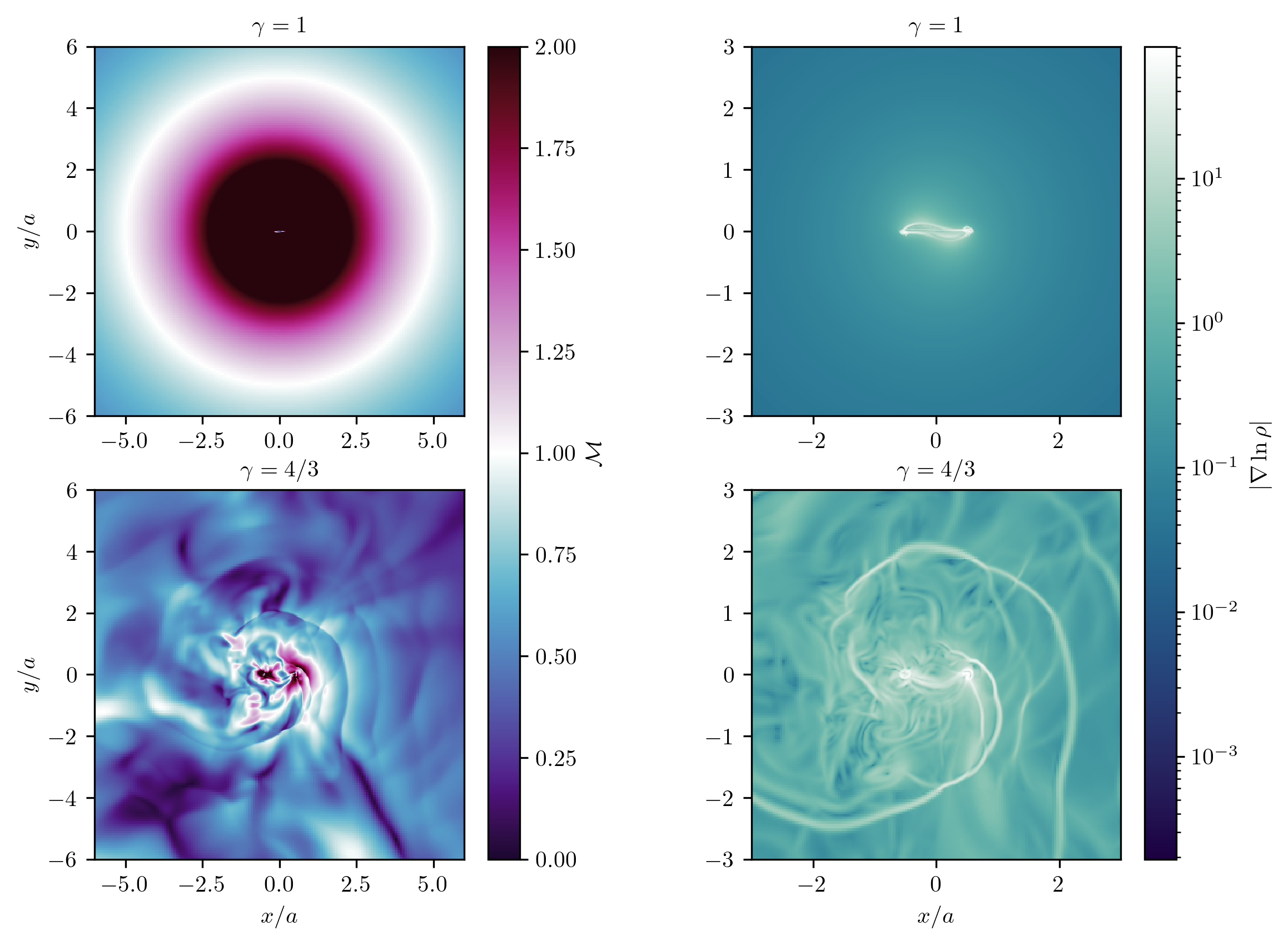}
    \caption{Mach number $\mathcal{M} = |\mathbf{v}|/c$ (left) and 
density gradient schlieren $|\nabla\ln\rho|$ (right) in 
the orbital plane at $\xi = 10$ at $t = 3t_B$ for $\gamma = 1$ (top) 
and $\gamma = 4/3$ (bottom). For $\gamma = 1$ the sonic 
surface ($\mathcal{M} = 1$, visible as the 
light-to-dark transition) is smooth, circular, and 
stable -- the Bondi saddle point is intact. For $\gamma 
= 4/3$ the saddle point has been annihilated by 
cumulative entropy injection: the sonic surface is 
disrupted and supersonic pockets fill the interior. The 
density gradient confirms the transition from laminar 
to fully turbulent flow.}
    \label{fig:iso_rel_zoom}
\end{figure*}

For $\gamma = 4/3$, the flow at $\xi = 0.1$ and $\xi = 1$
is organized and qualitatively similar to the isothermal case,
with coherent spiral structure in the corotating frame.
At $\xi = 10$ the flow is fully turbulent: the spiral
pattern has been destroyed, replaced by chaotic, multi-scale
density fluctuations filling the domain to $r \sim 30\,a$. 
At the orbital scale (Figure~\ref{fig:iso_rel_zoom}), 
the Mach number field reveals that the sonic surface has 
been completely disrupted for $\gamma = 4/3$: supersonic 
pockets fill the interior and two shocks tracing the $m 
= 2$ pattern at $r \sim 2a$ mark the remnant of the 
annihilated separatrix. Inside it, filaments, vortices, 
and secondary shocks fill the cavity.
% At the orbital scale (Figure~\ref{fig:iso_rel_zoom}), 
% two shocks tracing the $m = 2$ pattern at $r \sim 2a$ 
% marks the disrupted sonic surface; inside it, filaments, vortices, and secondary
% shocks fill the cavity. 
The transition between organized and turbulent flow occurs
between $\xi = 1$ and $\xi = 10$, consistent with the
predicted threshold $\xi_m = 4$ at which the sonic surface
first encloses the binary (Section~\ref{sec:entropy_source}).

Figure~\ref{fig:mon_schlieren} shows the same diagnostic
for $\gamma = 5/3$. The flow is disordered at all~$\xi$,
including $\xi = 0.1$ where the orbital Mach number is
subsonic. At $\xi = 10$ the turbulence fills the domain
out to $\sim 30\,a$ with no coherent large-scale structure.
The absence of a detached sonic surface  means there is no causal
membrane shielding the outer flow from the binary's
gravitational stirring: perturbations propagate outward
without impediment.
\begin{figure*}[t]
    \centering
    \includegraphics[width=\textwidth]{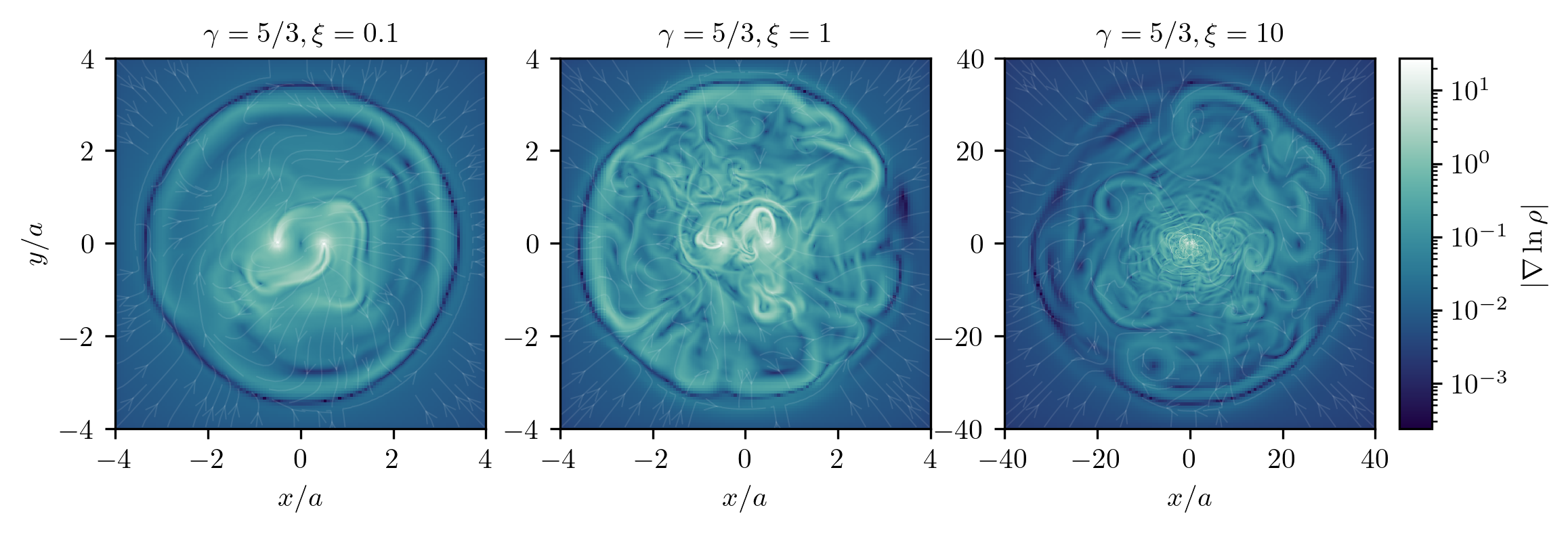}
    \caption{Density gradient schlieren $|\nabla \ln\rho|$ with velocity streamlines overlaid for
$\gamma = 5/3$ in the orbital plane at $t \geq 3\,t_B$. The
flow is disordered at all~$\xi$: no sonic surface exists to
shield the outer flow. Turbulent structure extends to
$\sim 30\,a$ at $\xi = 10$.}
    \label{fig:mon_schlieren}
\end{figure*}
\subsection{Radial flow structure}
\label{sec:profiles}

Figure~\ref{fig:profiles} shows time- and angle-averaged
radial profiles of the density~$\langle\rho\rangle$,
squared sound speed~$\langle c^2\rangle$, absolute radial
velocity~$\langle|v_r|\rangle$, and turbulent velocity
dispersion~$\langle\delta v\rangle$ at $\xi = 10$ for
each~$\gamma$. Profiles are averaged over spherical
shells centered on the binary center of mass for $r > a$,
and over the quasi-steady-state interval $t > 10$~orbits.
The velocity dispersion is computed via Reynolds
decomposition: for each cell, the time-mean velocity is
subtracted before computing the root-mean-square (RMS) fluctuation, so
$\delta v$ measures genuine temporal variability rather
than organized non-radial flow (e.g.\ the high-density
bridge in the isothermal case). Gray dotted lines show the
analytic Bondi solution for each~$\gamma$ (with the exception of the $\propto r^{-1}$ line shown for $\langle \rho \rangle$); light traces show the temporal variance of the angle-averaged profiles over the averaging interval of one orbit.

For $\gamma = 1$, all four profiles track the analytic Bondi solution reasonably well. The density follows $\rho \propto r^{-3/2}$ in the supersonic interior, the
sound speed is constant by construction, and the radial
velocity follows the freefall scaling
$|v_r| \propto r^{-1/2}$. The velocity dispersion is
negligible ($\delta v / |v_r| \lesssim 10^{-2}$),
confirming that the flow is laminar and steady.

For $\gamma = 4/3$, the flow roughly tracks the expected Bondi profile
even though the flow is turbulent. For $\gamma  = 5/3$, the profiles deviate from the Bondi solution inside $r \sim 5a$. The density flattens to
$\rho \propto r^{-1}$, shallower than the freefall
scaling. The ``temperature'' profiles of the adiabatic flows, by contrast, are
largely unaffected: the squared sound speed follows
$c^2 \propto r^{-3(\gamma - 1)/2}$ --  the canonical
Bondi scaling -- for both $\gamma = 4/3$ and $\gamma = 5/3$. 
The entropy instability redistributes gas spatially, flattening the density
profile for the stiffer gas ($\gamma = 5/3$), but preserves the temperature stratification set by the gravitational potential.

The velocity structure reveals the mechanism of
suppression. The radial infall velocity drops to one to
two orders of magnitude below the freefall value for both
adiabatic cases. The turbulent velocity dispersion,
measured via Reynolds decomposition, reaches
$\delta v \sim 0.1$--$1\,c_\infty$ in the interior,
exceeding the coherent radial infall by a factor of a
few to $\sim 10$ for $\gamma = 4/3$ and $\sim 10$--$100$
for $\gamma = 5/3$ in the interior ($r \lesssim 10a$). 
The flow is not accreting in any organized
sense: large-scale convective eddies circulate gas
through the Bondi sphere, and accretion proceeds as a
small net residual of nearly canceling inflow and
outflow. The temporal variance in $|v_r|$ (light traces) is
comparable to the mean itself for $\gamma > 1$, while
the variance in $\delta v$ is modest -- the turbulence
is statistically steady even as the instantaneous radial
velocity fluctuates wildly.

The $\rho \propto r^{-1}$ inner scaling for $\gamma = 5/3$ matches the
density profiles found in a number of previous studies, including turbulent non-radiative accretion onto a point mass in hydrodynamics and MHD \citep{Guo2023, Cho2023, Guo2025}.
%in previous studies \citep[see e.g.,][]{White+2020} 
%in magnetized Bondi accretion, where the flattening is
%driven by convective entropy transport. 
That the same
scaling emerges here -- from purely hydrodynamic,
binary-driven stirring rather than magnetic turbulence or externally imposed hydrodynamic turbulence
-- suggests the inner profile is set by the convective
equilibrium of the envelope rather than by the mechanism
that seeds the turbulence \citep[e.g.,][]{Ressler+2020}. This universality reinforces
the connection between the binary Bondi problem and the
broader convection-dominated accretion flow / radiatively inefficient accretion flow  literature: in both cases, the accretion rate at small radii is governed by the
turbulent transport properties of the envelope.
\begin{figure*}[t]
    \centering
    \includegraphics[]{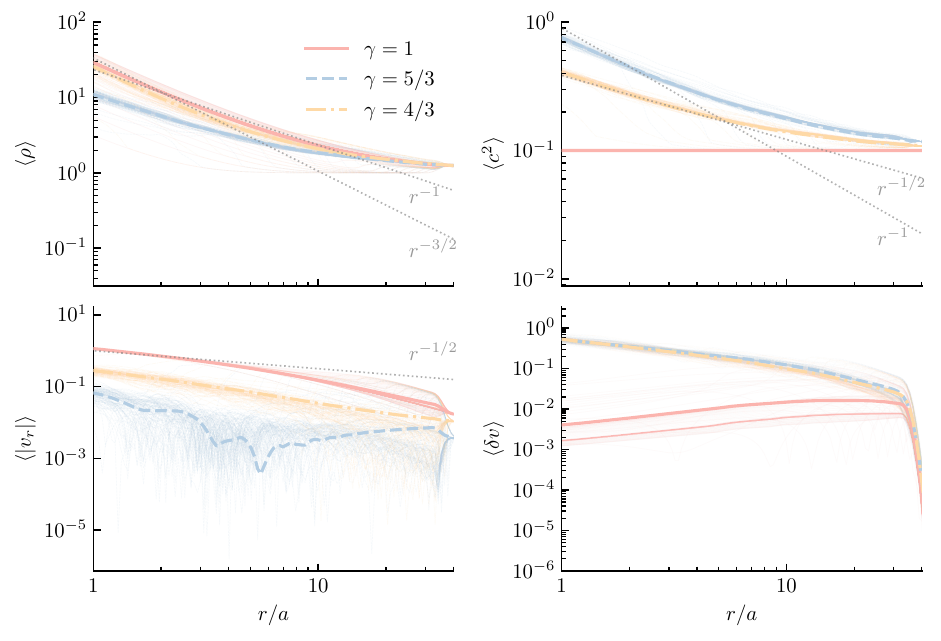}
    \caption{Time- and angle-averaged radial profiles at
$\xi = 10$: density (top left), squared sound speed (top
right), absolute radial velocity (bottom left), and
turbulent velocity dispersion via Reynolds decomposition
(bottom right). Gray dotted lines show the analytic Bondi
profiles; reference power laws are labeled. Lighter traces show the temporal variance of the angle-averaged profiles. For $\gamma = 1$ (solid, pink), all
profiles track the Bondi solution and $\delta v$ is
negligible: the flow is laminar. For $\gamma = 4/3$ (dash-dotted yellow), the density profile obeys the Bondi character reasonably well, but the $\gamma = 5/3$ flattens towards $\rho \propto r^{-1}$ inside $r \sim 5a$ while the temperature profiles $c^2 \propto r^{-3(\gamma-1)/2}$ for both are preserved. The radial velocity is suppressed by factors of a few below freefall; the turbulent dispersion exceeds it by a factor of a few to $\sim10$, confirming that the interior is a turbulence-dominated 
reservoir with accretion as a residual.}
    \label{fig:profiles}
\end{figure*}
\subsection{Turbulence Character}
\label{sec:turbulence_character}

We characterize the irreversible entropy field via the 
angular power spectrum of $\kappa \equiv P/\rho^\gamma$. 
Since $\kappa$ is an adiabatic invariant -- constant along 
streamlines in the absence of shocks -- all spatial 
structure in $\kappa$ traces entropy generation rather than 
organized flow. At each radius $r$, we interpolate $\kappa$ 
onto a regular equirectangular grid in $(\theta, \phi)$, 
subtract the shell mean, and compute the angular power 
spectrum via 2D fast Fourier transform (FFT) with $\sin\theta$ quadrature weighting:
\begin{equation}
    C_\ell^{(\kappa)}(r) 
    = \frac{1}{2\ell + 1}
      \sum_{m=-\ell}^{\ell} 
      \left| \hat{\kappa}_{\ell m}(r) \right|^2\,,
    \label{eq:Cell}
\end{equation}
where $\hat{\kappa}_{\ell m}$ are the 2D Fourier 
coefficients of the weighted fluctuation field 
$\delta\kappa\,\sin\theta$, and the physical wavenumber is 
$k = \ell/r$ in units of ${\rm rad}/a$. The 2D FFT on an 
equirectangular grid approximates a true spherical harmonic 
transform; the approximation is adequate for power-law 
slope estimation but not for absolute amplitudes at low 
$\ell$. Spectra are averaged over seven logarithmically spaced 
shells spanning $r/a \in [1, 10]$, interior to the Bondi 
radius $R_B = 10\,a$.

For $\gamma = 1$, the spectrum sits at machine precision 
($C_\ell \sim 10^{-38}$) at all scales, confirming that 
the isothermal flow generates no entropy despite the 
presence of shocks and spiral structure. This null test 
establishes that the signal in the adiabatic cases reflects 
genuine entropy fluctuations rather than numerical or 
geometric artifacts.

For $\gamma = 4/3$ and $5/3$, the spectra are elevated by 
orders of magnitude. The $\gamma = 5/3$ spectrum carries 
more power than $\gamma = 4/3$ at all scales, consistent 
with more vigorous large-scale entropy rearrangement driven 
by convection in the neutrally stratified flow. Both 
adiabatic spectra follow $k^{-3.5 \pm 0.1}$ over the 
resolved range $k \sim 0.5$--$6\,{\rm rad}/a$. 
This slope is steeper than standard cascade predictions
(e.g., $k^{-5/3}$ Kolmogorov or $k^{-2}$ Burgers), 
consistent with a spectrum dominated by isolated coherent structures 
rather than a volume-filling cascade. It is consistent with the 
coherent structures visible in the density gradient images 
(Figures~\ref{fig:iso_rel_schlieren}--\ref{fig:mon_schlieren}).

The two adiabatic cases have distinct physical origins 
despite their similar spectral slopes. For $\gamma = 4/3$, 
the causal chain runs from the supersonic propeller to 
standing orbital shocks, to entropy accumulation, to 
breakout: shocks \emph{cause} the instability. For $\gamma 
= 5/3$, the propeller is marginally sonic ($\mathcal{M}_p 
= 1$) and generates no standing shocks in the corotating 
frame; the instability is convective, driven by entropy 
rearrangement in a neutrally stratified flow. The resulting 
convective velocities are of order $v_{\rm orb}$ in a gas 
where $v/c \sim 1$ at all radii, so the turbulent motions 
produce shocks in the saturated state. The similar 
$k^{-7/2}$ slopes reflect similar shock front morphology 
in the saturated state, not a common driving mechanism.

The absence of a developed inertial-range cascade below 
$k_a$ reflects the nature of the entropy source. The 
$\kappa$ field on each shell is dominated by a small 
number of shock fronts and convective plume boundaries 
rather than a space-filling turbulent cascade; power is 
injected at $k \sim k_a$ by the binary and dissipated 
at the grid scale without an extended transfer between 
scales. The $k^{-7/2}$ slope is therefore set by the 
morphology of these coherent structures -- finite-width 
shock fronts in a smooth background -- rather than by, say,  a 
Richardson-type cascade. With regard to resolution, our 
convergence tests (Section~\ref{sec:convergence}) 
confirm that $\eta$ is converged to $<5\%$, but we have 
not performed systematic resolution studies of the 
spectral slope. Higher resolution would push $\ell_{\rm 
max}$ upward and extend the resolved range; we leave 
this to future work.

\subsection{Temporal Variability}
\label{sec:variability}
\begin{figure}[t]
    \centering
    \includegraphics[width=\columnwidth]{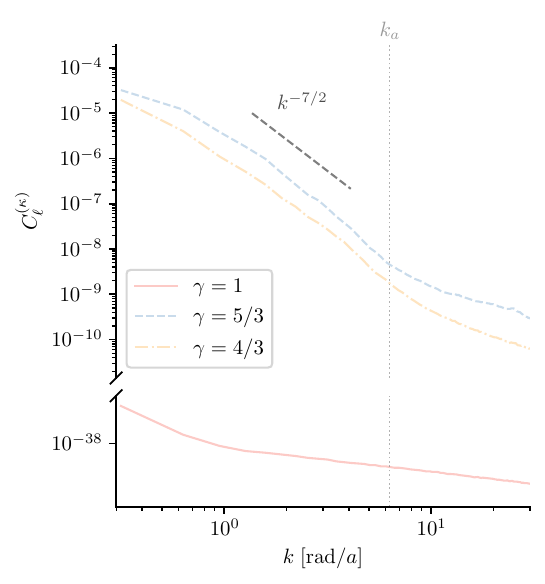}
    \caption{Angular power spectrum of the entropy proxy $\kappa = P/\rho^\gamma$ at $\xi = 10$. For each of seven logarithmically spaced shells in $r/a \in [1, 10]$, the fluctuation field $\delta\kappa = \kappa - \langle\kappa\rangle_{\rm shell}$ is interpolated onto an equirectangular $(\theta,\phi)$ grid and decomposed via 2D FFT with $\sin\theta$ weighting; the resulting $C_\ell$ are averaged across shells and mapped to spatial wavenumber $k = \ell/r$. The isothermal case ($\gamma = 1$) returns $C_\ell \sim 10^{-38}$ at all scales, confirming that the adiabatic spectra reflect genuine entropy production rather than numerical or geometric artifacts. Both $\gamma = 5/3$ and $\gamma = 4/3$ follow $k^{-7/2}$ over the resolved range $k \lesssim k_a$, steeper than Kolmogorov ($k^{-5/3}$) or Burgers ($k^{-2}$) scaling, consistent with a spectrum set by isolated shock fronts and convective boundaries rather than a volume-filling cascade. The $\gamma = 5/3$ spectrum carries more power at all scales, reflecting larger per-shock entropy jumps in the stiffer equation of state. The upturn above $k_a$ is a pixel window artifact of the equirectangular interpolation and should not be interpreted as physical.}
    \label{fig:entropy_spectrum}
\end{figure}
Figure~\ref{fig:accretion_power} shows Lomb--Scargle
periodograms of the total accretion rate
$\dot{M}_{\rm tot}(t)$ (dark) and the individual component
rates $\dot{M}_{1,2}(t)$ (light) at $\xi = 10$ for
each~$\gamma$. Since in practice the emission from each
component can dominate a distinct spatial region, the
individual periodograms can also be an observationally
relevant quantity.

For $\gamma = 5/3$ (top), both the total and individual
periodograms are pure broadband noise with no coherent
feature at any orbital harmonic. The turbulence has
completely erased the orbital signal. For $\gamma = 4/3$ 
(middle), a residual peak at $2\Omega$ rises $\sim 
5$--$10\times$ above the turbulent continuum in both the 
total and individual rates. The $2\Omega$ dominance 
reflects the equal-mass symmetry: the $q = 1$ potential 
repeats every half orbit. The orbital signature survives 
but is partially buried in turbulent noise. For $\gamma = 
1$ (bottom), both the individual component rates and the 
total show coherent peaks at $4\Omega$ and $8\Omega$, with 
no significant power at $2\Omega$. The accretion rate 
responds nonlinearly to the $2\Omega$ quadrupolar forcing, 
transferring power to the first overtone and its harmonic. 
An observer would detect periodicity at $4\Omega$ and 
infer half the true orbital period.

Moreover, we emphasize that the 
suppression of $2\Omega$ and the dominance of 
$4\Omega$ and $8\Omega$ follows from the quadratic 
nonlinearities of the Euler equations. The stable sonic 
surface screens the direct $2\Omega$ forcing from the 
binary quadrupole; the residual signal that reaches the 
sinks is shaped by nonlinear mode coupling. At second 
order in the perturbation amplitude, products of 
first-order quantities oscillating at $2\Omega$ generate 
forcing terms at $4\Omega$ via $\cos^2(2\Omega t) = 
\frac{1}{2}(1 + \cos(4\Omega t))$, with both the 
advective term $\delta\mathbf{v}\cdot\nabla\delta\mathbf{v}$ 
and the nonlinear pressure response contributing. The 
$8\Omega$ signal appears at third order by the same 
mechanism.

Additionally, the broadband continuum in the
$\gamma = 5/3$ and $\gamma = 4/3$ periodograms shares a
similar red-noise slope, consistent with the same
$k^{-7/2}$ spatial turbulent spectrum imprinting on the
temporal variability -- eddies sweeping past the accretion
sphere map the spatial slope to a temporal one via Taylor's
frozen-flow hypothesis. The $2\Omega$ orbital signal
survives above the continuum for $\gamma = 4/3$ but falls
below the detection threshold for $\gamma = 5/3$. However,
a phase-coherent folding analysis -- averaging
$\dot{M}(t)$ over ${\sim}100$ orbits as a function of
orbital phase -- reveals a coherent $2\Omega$ modulation
in both adiabatic cases, with comparable fractional
amplitudes of ${\sim}1$--$3\%$. The periodogram
non-detection for $\gamma = 5/3$ is therefore a
sensitivity limitation rather than a physical absence: the
coherent signal has the same strength, but the broadband
turbulent noise floor is proportionally higher, pushing the
peak below the false-alarm threshold. This confirms that
the $2\Omega$ gravitational forcing of the binary is
universal across equations of state, and that the dominant
source of temporal variability in all cases is the
turbulent cascade rather than the orbital dynamics.

The broadband floor in the $\gamma = 1$ panel is elevated 
relative to the adiabatic cases and is a numerical 
artifact: the isothermal flow concentrates mass into a 
narrow high-density bridge between the two accretors, and 
the finite accretion kernel clips this filament differently 
at each timestep, producing rapid oscillations in the 
instantaneous $\dot{M}$ at frequencies well above the 
orbital harmonics. For $\gamma = 4/3$ and $\gamma = 5/3$, 
turbulence broadens density gradients over many cells and 
the kernel integral is smooth, suppressing this artifact 
despite the stronger physical variability in those cases. 
The relevant diagnostic is therefore peak height above the 
noise floor and not absolute power. By this measure the 
coherent peaks at $4\Omega$ and $8\Omega$ in the $\gamma = 
1$ case rise $1$--$1.5$ decades above the artifact floor 
and are robust.

The progression from broadband noise ($\gamma = 5/3$)
through marginal detection ($\gamma = 4/3$) to coherent
periodicity ($\gamma = 1$) has direct implications for
binary AGN searches: the thermodynamic regime of the
ambient gas determines not only the accretion efficiency
but also the detectability of the orbital signal.

\subsection{Accretion Efficiency}
\label{sec:efficiency}

Figure~\ref{fig:eta_vs_xi} shows the accretion efficiency
$\eta \equiv \dot{M}_{\rm binary}/\dot{M}_{\rm Bondi}$
as a function of~$\xi$ for each~$\gamma$. Error bars
span the 10th--90th percentile of $\eta(t)$. 
For $\gamma = 1$, $\eta$ rises monotonically 
from $\approx 0.5$ at $\xi = 0.1$ to $\approx 1$ at
$\xi = 10$. At low $\xi$ the components accrete
independently, recovering the predicted
$\eta = 1/2$ for equal masses
(Eq.~\ref{eq:eta_independent}). At high $\xi$ the
bridge channels mass cooperatively and
$\eta \to 1$.

For $\gamma = 4/3$, $\eta$ tracks the independent rate
at low $\xi$, rising from $\approx 0.42$ at $\xi = 0.1$
to $\approx 0.51$ at $\xi = 1$. Beyond the container
threshold $\xi_m = 4$, $\eta$ drops sharply to
$\approx 0.32$ at $\xi = 10$. The turbulence does not
merely prevent cooperation -- it actively suppresses
accretion below the independent rate. The overpressured,
entropy-laden gas resists infall more effectively than
the undisturbed Bondi profile.

Both $\gamma = 1$ and $\gamma = 4/3$ flows develop spiral shocks at $\xi = 1$, yet the
isothermal efficiency already exceeds $1/2$ while the
polytropic efficiency does not. The distinction lies in
what the shock does to the gas. An isothermal shock
dissipates kinetic energy that is radiated away
instantaneously: the gas decelerates without pressurizing,
making it easier to capture. The spiral arms actively
assist accretion by removing the kinetic energy that would
otherwise carry gas past the accretor. An adiabatic shock
converts kinetic energy into thermal energy: the gas
decelerates but heats up, and the resulting pressure
cushion resists further infall. The funneling effect of
the spiral arms and the post-shock pressurization nearly
cancel, leaving $\eta \approx 1/2$.

For $\gamma = 5/3$, $\eta$ is suppressed at all~$\xi$,
ranging from $\approx 0.27$ at $\xi = 0.1$ to
$\approx 0.10$--$0.12$ at $\xi \geq 1$. The error bars
are small, reflecting saturated turbulence with steady
suppression rather than organized flow. The absence of a
threshold in~$\xi$ is consistent with the geometric
instability of the attached sonic surface.
\begin{figure}[t]
    \centering
    % First Subfigure
    \begin{subfigure}[b]{\columnwidth}
        \centering
        \includegraphics[width=\linewidth]{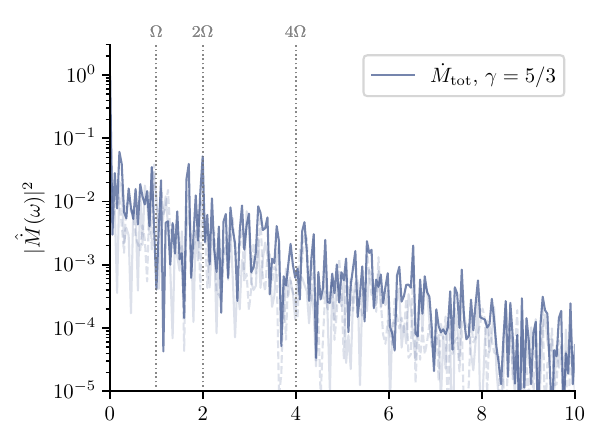}
    \end{subfigure}
    % \\[2ex] % Adds a specific amount of vertical gap

    % Second Subfigure
    \begin{subfigure}[b]{\columnwidth}
        \centering
        \includegraphics[width=\linewidth]{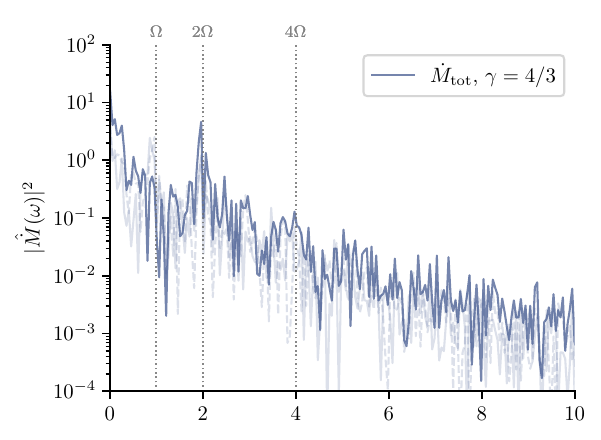}
    \end{subfigure}
    % \\[2ex]

    % Third Subfigure
    \begin{subfigure}[b]{\columnwidth}
        \centering
        \includegraphics[width=\linewidth]{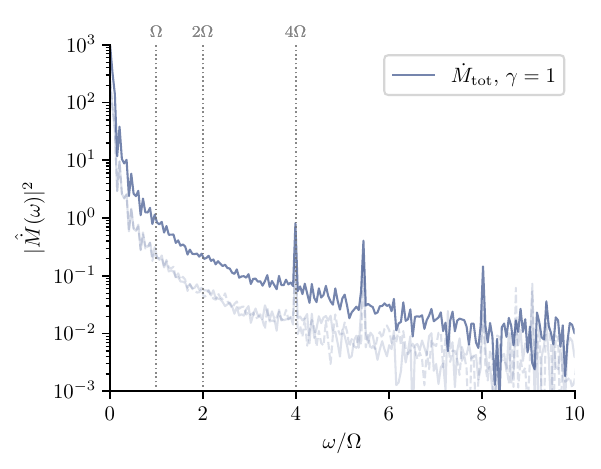}
    \end{subfigure}
    \caption{Lomb--Scargle periodograms of $\dot{M}_{\rm tot}(t)$ (dark) and individual component rates (light) at $\xi = 10$. For $\gamma = 5/3$, the orbital signal is fully buried in broadband turbulent noise. For $\gamma = 4/3$, a residual peak at $2\Omega$ survives above the continuum. For $\gamma = 1$, coherent peaks appear at $4\Omega$ and $8\Omega$, with no power at $2\Omega$: the accretion rate responds nonlinearly to the quadrupolar forcing, transferring power to overtones. An observer would infer half the true orbital period.}
    \label{fig:accretion_power}
\end{figure}
%
%%%%%%%%%%%%%%%%%%%%%%%%%%%%%%%%%%%%%%%%%%%%%%%%%%%%%%%%%%%

\section{Discussion}
\label{sec:discussion}

An embedded binary in adiabatic gas is self-limiting: the
orbit that gravitationally drives accretion simultaneously generates the
entropy that disrupts it. We discuss the implications for
accretion physics, binary detection, and specific
astrophysical systems.

\subsection{A Third Route to Convection-Dominated Accretion}
\label{sec:cdaf}

Two routes to convectively unstable accretion are known: dissipation of rotational energy in rotating advection-dominated flows
\citep{Narayan+Yi+1994,Narayan+Yi+1995}, potentially producing the CDAF
regime \citep{Stone+Pringle+Begelman+1999, Igumenschev+Abramowicz+2000, Narayan+2000, Quataert+Gruzinov+2000}, and the dissipation of magnetic energy in magnetized accretion from a uniform medium \citep{Igumenshchev2002}.    

Binary Bondi accretion provides a third route. The rotating
quadrupolar potential drives shocks that generate entropy
without viscosity, magnetic fields, or angular momentum in
the ambient medium. The resulting flow shares the essential
features of CDAFs -- entropy-driven convection opposing
infall, convective recycling, and suppression of $\dot{M}$
-- but the energy source is the orbital rotational energy rather than `viscous'
dissipation of the gas rotational energy.
A related phenomenon appears in circumbinary
disks, where \citet{Ryan+MacFadyen+2017} showed that tidally
excited spiral shocks provide an effective viscosity
$\alpha \sim 0.01$ through purely hydrodynamic dissipation.

The suppression of $\eta$ below unity reflects convective recycling rather than 
unbound mass loss: the turbulent interior circulates gas through
large-scale eddies, and the accretion proceeds as a small net residual of nearly canceling inflow and outflow motions (Section~\ref{sec:profiles}). Whether a fraction 
of this recycled gas ultimately escapes to infinity with positive 
Bernoulli parameter -- as in the ADIOS picture \citep{Blandford+Begelman+1999} -- is not determined by the present simulations, which maintain a fixed mass supply at the outer boundary. 

The spectral distinction between the $k^{-7/2}$ slope found here and the $k^{-5/3}$ Kolmogorov scaling expected for vortical turbulence is a testable prediction for future MHD simulations.

\begin{figure}[t]
    \centering
    \includegraphics[width=\columnwidth]{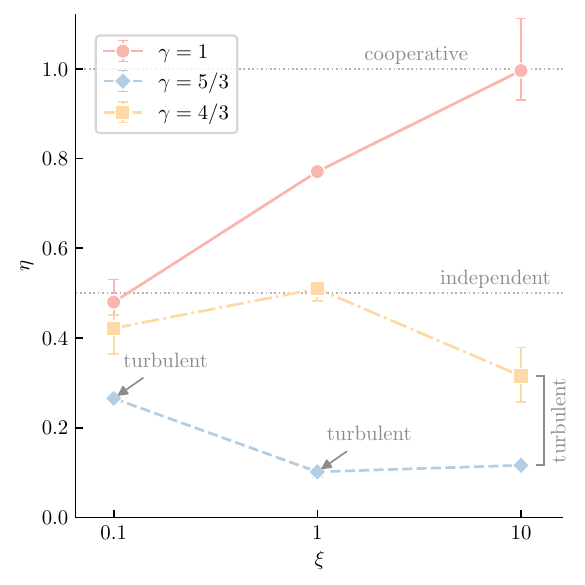}
    \caption{Accretion efficiency
    $\eta \equiv \dot{M}_{\rm binary}/\dot{M}_{\rm Bondi}$ versus
    compactness $\xi$ for equal-mass binaries at three adiabatic
    indices. The colors and line styles are the same as in Figure \ref{fig:entropy_spectrum}. Error bars span the 10th--90th percentile of $\eta(t)$ measured over 5-orbit windows after $t > \,t_B$.
    Isothermal gas ($\gamma = 1$) rises monotonically to
    $\eta \to 1$: the binary accretes cooperatively through a
    high-density bridge. Relativistic gas ($\gamma = 4/3$) tracks
    the independent rate ($\eta = 1/2$) at low $\xi$, then drops
    sharply once the sonic surface encloses the binary
    ($\xi_m = 4$), as shock-generated entropy drives convective
    turbulence. Monatomic gas ($\gamma = 5/3$) is turbulent and
    suppressed at all~$\xi$. Dotted lines mark the cooperative
    ($\eta = 1$) and independent ($\eta = 1/2$) limits.}
    \label{fig:eta_vs_xi}
\end{figure}

\subsection{Where Are the Periodic Binaries?}
\label{sec:detectability}

Searches for sub-parsec SMBH binaries have targeted periodic
signatures in AGN light curves
\citep{Graham+2015,Charisi+2016,Liu+2019}, motivated by
circumbinary disk simulations predicting periodic accretion
modulation \citep[e.g.,][]{MacFadyen+Milosavljevic+2008,
DOrazio+Haiman+MacFadyen+2013,Duffell+2020}. The yield has
been disappointing: \citet{Liu+2019} found most Pan-STARRS1
candidates did not persist, \citet{Vaughan+2016} argued
PG~1302$-$102 is consistent with red noise, and
\citet{ElBadry+2025} showed that all 181 Gaia~DR3 periodic
AGN candidates were false positives.

Our results offer a natural explanation. In radiatively
inefficient environments ($t_{\rm cool} \gg NT$), the entropy
instability drives turbulence that partially ($\gamma = 4/3$)
or fully ($\gamma = 5/3$) buries the orbital signal in
broadband noise (Section~\ref{sec:variability}). In
radiatively efficient environments ($t_{\rm cool} \ll NT$),
the binary does produce coherent periodicity -- but at
$4\Omega$ and $8\Omega$, not at the expected $2\Omega$. The
accretion rate responds nonlinearly to the quadrupolar
forcing, transferring power to overtones. An observer
%searching at $\Omega$ or $2\Omega$ would detect nothing;
%one
who found the $4\Omega$ peak would infer half the true
orbital period. The periodicity channel is not closed, but
it is tuned to an unexpected frequency.

Three observational implications follow. First, the absence
of periodicity at $2\Omega$ does not exclude a binary:
adiabatic binaries erase the signal, and isothermal binaries
shift it. 
%Future surveys should extend periodicity searches
%to $4\Omega$, corresponding to half the expected orbital
%period. 
Second, the ratio of power at $2\Omega$ to $4\Omega$
encodes the mass ratio: for $q = 1$ the half-orbit symmetry
suppresses $2\Omega$ in the total accretion rate.
% for $q \neq 1$ the symmetry breaks and $2\Omega$ should reappear.
Third, kinematic signatures such as Doppler boosting of
relativistic jets \citep{ONeill+2022,Kiehlmann+2025} bypass
the accretion flow entirely and are unaffected by either
the entropy instability or the nonlinear frequency shift.
These remain the most robust electromagnetic probe of
embedded binaries.

\subsection{Astrophysical Applications}
\label{sec:applications}

The single comparison $t_{\rm cool}$ versus $NT$ partitions 
astrophysical binaries into three regimes. The thermodynamic 
regime -- adiabatic, marginal, or isothermal -- depends on 
the cooling time at $r \sim a$, set by local microphysics 
(density, temperature, opacity), and is insensitive to the 
global flow geometry. The predicted efficiency $\eta$, by 
contrast, is calibrated to the spherical Bondi geometry 
simulated here and applies directly only when the flow at 
$r \sim a$ is pressure-supported rather than rotationally 
supported.

\subsubsection{Adiabatic regime: $t_{\rm cool} \gg NT$}

Hot, radiatively inefficient environments place binaries 
in the adiabatic regime. Post-merger galaxies hosting SMBH 
binaries in tenuous plasma at temperature $\Theta \sim 10^7\;\text{K}$, 
and compact object binaries orbiting inside common stellar 
envelopes at $\Theta \sim 10^5\;\text{K}$, both satisfy 
$t_{\rm cool}/NT \gg 1$ across a wide range of separations 
and masses. In both cases the entropy instability is active, 
suppressing accretion to $\eta \sim \mathcal{O}(0.1)$. For SMBH 
binaries, laminar Bondi-based estimates of gas hardening 
adopted in population synthesis models 
\citep{Haiman+2009,Sesana+2013,Kelley+2017a} overestimate 
the hardening rate by $1/\eta \sim 3$--$10$, providing a 
thermodynamic ingredient in the final parsec problem 
\citep{Begelman+Blandford+Rees+1980,
Milosavljevic+Merritt+2003} that is independent of 
feedback. For common envelope systems, Bondi--Hoyle--Lyttleton 
prescriptions used in single-body drag estimates 
\citep{Macleod+RamirezRuiz+2015a,MacLeod+2017,De+2020} do 
not capture the entropy feedback from the binary's orbital 
motion and likely overestimate the accretion rate onto the 
embedded pair.

In both cases the principal caveat is geometry: the giant 
envelope has a steep density gradient and the SMBH 
environment may carry significant angular momentum, both 
of which violate the uniform pressure-supported medium 
assumed here. The thermodynamic regime ($t_{\rm cool}/NT 
\gg 1$) is robust because it depends only on local 
microphysics at $r \sim a$, but the quantitative $\eta$ 
should be taken as indicative.

\subsubsection{Marginal regime: $t_{\rm cool} \sim NT$}

Dense, optically thick environments such as AGN disks place 
binaries near marginal cooling \citep{Bartos+2017,Stone+2017,
Tagawa+2020}. The thermodynamic outcome is sensitive to 
local conditions and the quantitative $\eta$ from our 
spherical simulations does not apply directly: the disk 
scale height is comparable to or smaller than $R_B$, 
Keplerian shear introduces a velocity gradient across the 
Hill sphere, and residual angular momentum produces a 
circumbinary mini-disk. In the rotationally supported 
limit the propeller Mach number vanishes and the entropy 
instability is inoperative; a distinct suppression 
mechanism driven by the narrowing of accretion streams 
operates instead \citep{Tiede+2025}.

\subsubsection{Isothermal regime: $t_{\rm cool} \ll NT$}

Efficiently cooled environments -- molecular clouds, atomic 
cooling halos, and globular cluster gas -- place binaries 
in the isothermal regime regardless of their microscopic 
$\gamma$. Post-shock entropy is radiated before the binary 
completes a single orbit, and the gas behaves as effectively 
isothermal. Binaries in these environments accrete 
cooperatively through a high-density bridge, with $\eta 
\to 1$ and mass growth rates approaching twice the 
independent-accretor value. These systems are efficient 
accretors but poor periodic sources: the isothermal flow 
concentrates orbital power into overtones at $4\Omega$ and 
$8\Omega$ rather than $2\Omega$ 
(Section~\ref{sec:variability}), and kinematic signatures 
such as Doppler boosting \citep{ONeill+2022} remain the 
most robust electromagnetic probe.

\subsection{The Role of Cooling}
\label{sec:cooling}

Our simulations treat two limits: $\gamma = 1$ (instantaneous
cooling) and $\gamma > 1$ (no cooling). The transition is
sharp because the breakout number $N$ is only
$\mathcal{O}(1)$--$\mathcal{O}(10)$: a system either cools faster than $NT$
or it does not. When the cooling luminosity approaches
$L_{\rm Edd}$, radiation pressure stiffens the effective
equation of state toward $\gamma = 4/3$ -- precisely the
regime where the entropy instability is most potent.
Eddington-limited accretion onto a single object proceeds
unimpeded; onto a binary, it triggers the instability.

\citet{Tiede+DOrazio+2025} reach a consistent conclusion from
the spectral energy distribution (SED) side: only emission mechanisms with cooling times
shorter than $T$ carry orbital-period variability, while
slower channels are damped. Our simulations provide the
hydrodynamic mechanism: entropy-driven convection scrambles
the accretion modulation before it imprints on the luminosity.

\subsection{Caveats}
\label{sec:caveats}

The orbit is held fixed; for widely separated systems the
inspiral timescale vastly exceeds the breakout timescale, so
the steady state is established before the separation changes
appreciably. We consider only $q = 1$; for $q \ll 1$ the entropy
source weakens as $q/(1+q)^2$ and the disruption
timescale increases, though the container condition
$\xi > \xi_m$ is unchanged. Magnetic fields are absent; MHD
effects could stabilize (tension) or enhance (magneto-rotational instability) the
turbulence, and the spectral prediction $k^{-7/2}$ versus
$k^{-5/3}$ provides a diagnostic.  More generally, hotter environments that most favor the entropy instability thermodynamically ($\gamma_{\rm eff}$ further from unity) also shrink the Bondi radius, reducing~$\xi$ and potentially moving the
system out of the deeply embedded regime. This
$\Theta$--$\xi$ anticorrelation is a natural self-regulation:
the conditions that make the instability most potent are
also the conditions where embedding is hardest to achieve.
% The common envelope entry in
% Table~\ref{tab:applications} illustrates this tension
% explicitly. 
The ambient medium is at rest and carries no angular momentum; a nonzero bulk velocity introduces BHL effects \citep{Edgar+2004,Antoni+2019}, and
residual rotation would produce a circumbinary structure whose
interaction with the entropy instability is unexplored.

\section{Concluding Remarks}
\label{sec:conclusions}

A binary embedded in adiabatic gas generates the entropy that
throttles its own accretion. The threshold
$\xi_m = 4/(5 - 3\gamma) $ -- where $\xi \equiv R_B / a$ is the dimensionless number characterizing the total Bondi radius, $R_B$, of the binary relative to the binary separation, $a$ -- and the breakout timescale
$N \propto (\gamma - 1)^{-1}\sqrt{\xi}$ -- the number
of orbits required to annihilate the Bondi saddle point and
disrupt the transonic accretion flow -- are derived from first
principles; three-dimensional simulations confirm both.
Embedded binaries in hot, radiatively inefficient environments
are poor accretors (i.e., the ratio of the binary accretion rate relative to the
accretion rate of a single body of equal mass, $\eta \ll 1$) and poor periodic sources: gas-assisted hardening estimates for SMBH binaries in hot
atmospheres that adopt laminar Bondi rates will overpredict
the hardening rate by $1/\eta \sim 3$--$10$. Binaries in efficiently cooled environments accrete cooperatively and are undetectable through accretion-rate
variability; kinematic signatures such as Doppler boosting remain viable. The
dividing line is $t_{\rm cool}$ versus $N\,T$: systems that
cool faster than they heat remain organized; systems that do
not, go turbulent. This single comparison governs whether an
embedded binary grows, stalls, or destroys its own fuel
supply.

\begin{acknowledgements}
    We thank the anonymous referee for their useful comments that helped to improve this work. M.D. thanks Jenny Greene, Jonathan Zrake, Andrei Gruzinov, Andrew MacFadyen, Ilya Mandel, Yacine Ali-Ha{\"i}moud, Romain Teyssier, Jeremy Goodman, Tamar Faran, Yuri Levin, and Philip Jon {\O}stergaard Kirkeberg for very fruitful discussions surrounding this problem. 
\end{acknowledgements}

\software{\texttt{matplotlib} \citep{Hunter+2007}, \texttt{cmasher} \citep{vanDerVelden+2020}, \texttt{SIMBI} \citep{Dupont+2023}, \texttt{numpy} \citep{Harris+2020}, \texttt{scipy} \citep{Scipy+2020}}

%% For this sample we use BibTeX plus aasjournalv7.bst to generate the
%% the bibliography. The sample7.bib file was populated from ADS. To
%% get the citations to show in the compiled file do the following:
%%
%% pdflatex sample7.tex
%% bibtext sample7
%% pdflatex sample7.tex
%% pdflatex sample7.tex

\bibliography{refs}{}
\bibliographystyle{aasjournalv7}

%% This command is needed to show the entire author+affiliation list when
%% the collaboration and author truncation commands are used.  It has to
%% go at the end of the manuscript.
%\allauthors

%% Include this line if you are using the \added, \replaced, \deleted
%% commands to see a summary list of all changes at the end of the article.
%\listofchanges

\end{document}